\documentclass{nature}[a4paper]

\usepackage{bm}
\usepackage{authblk}
\usepackage{amsmath}
\usepackage{graphicx}
\usepackage{xcolor}
\usepackage[section]{placeins}
\makeatletter
\let\saved@includegraphics\includegraphics
\AtBeginDocument{\let\includegraphics\saved@includegraphics}
\renewenvironment*{figure}{\@float{figure}}{\end@float}
\makeatother
\usepackage[labelfont=bf]{caption}
\usepackage{cite}
\usepackage{soul}
\DeclareUnicodeCharacter{FF0C}{,}
\newcommand{\upcite}[1]{\textsuperscript{\scalebox{0.7}{\cite{#1}}}}

\title{Observation of ferron transport in ferroelectrics%Ferron Enables Dynamical Magnetoelectric Coupling%Transmission of Spin Information via spin-neutral Quasiparticle: Ferron%Ferron enables long range spin transport at room temperature
\footnotetext{$^\ast$  E-mail: chenxianzhe@fudan.edu.cn
\\}}

\author[1$\dagger$]{Kaiwen Shen}
\author[2$\dagger$]{Ping Tang}
\author[1$\ast$]{Xianzhe Chen}
\author[1]{Yifan Gao}
\author[3]{Yuanfei Fan}
\author[4]{Zejing Guo}
\author[1]{Yingfen Wei}
\author[1]{Hao Jiang}
\author[1]{Xumeng Zhang}
\author[1]{Ming Wang}
\author[4]{Pan He}
\author[4]{Wu Shi}
\author[2]{Jiahao Han}
\author[3]{Yizheng Wu}
\author[3,4]{Jian Shen}
\author[1]{Qi Liu}
\author[2,5,6]{Gerrit E.W. Bauer}
\author[1]{Ming Liu}

\affil[1]{Frontier Institute of Chip and System, State Key Laboratory of Integrated Chips and Systems, Zhangjiang Fudan International Innovation Center, Fudan University, Shanghai, 200433, China}
\affil[2]{WPI-AIMR, Tohoku University, 2-1-1 Katahira, 980-8577 Sendai, Japan}
\affil[3]{State Key Laboratory of Surface Physics, Department of Physics, Key Laboratory of Micro and Nano Photonic Structures (MOE), Fudan University, Shanghai 200433, China}
\affil[4]{State Key Laboratory of Surface Physics, Institute for Nanoelectronic Devices and Quantum Computing, Fudan University, Shanghai 200433, China}
\affil[5]{Kavli Institute for Theoretical Sciences, University of the Chinese Academy of Sciences, Beijing, 100190, China}
\affil[6]{Institute for Materials Research and CSRN, Tohoku University, 2-1-1 Katahira, 980-8577 Sendai, Japan}

\graphicspath{{figures/}}

\date{}                   
\begin{document}

    \maketitle

    \vspace{10pt}
    %\newpage
    \begin{abstract}
\textbf{
Ferroelectrics feature spontaneous electric dipolar order reconfigurable via electric fields\upcite{wei2018,wang2023,meier2012}.}\textbf{ %In ferroelectrics, an applied electric field induces reconfigurable spontaneous polarization: this hysteretic switching behavior is inherently governed by dipole correlation lengths, and thus enables long-range signal propagation mechanisms distinct from conventional charge-carrier transport
}%Theoretical studies have proposed "ferron" quasiparticles—collective excitations of this dipolar order—as potential analogs to magnons of magnetic materials for next-generation information and thermal management technologies
Recent theoretical studies of the collective excitations of this electric dipolar order give rise to the hope that ``ferron" quasiparticles may complement the magnons of magnetic materials in information and heat management technologies\upcite{tang2022,tang2022a,tang2023,tang2024,bauer2021,bauer2022,bauer2023}. %%add reference
 {%However, direct experimental evidence of ferron transport is still lacking.}
 Yet direct experimental evidence of ferron transport remains elusive. Here we demonstrate efficient ferron injection and detection enabled by ferromagnetic metal contacts, achieving nonlocal signal transmission over micrometer distances in a prototypical ferroelectric PMN-PT. %Here we observe efficient ferron injection and detection {%while the combined spin Hall effect and spin-dependent conductivities of 
 %via ferromagnetic metal contacts}. We observe non-local signal transmission through a typical ferroelectric over micrometers. 
 The transmission efficiency can be switched by external magnetic fields that couple to the contacts and gate electric fields that control the ferron excitations. Ferron-based devices open new power saving strategies that employ ferroelectric materials in a future sustainable information society\upcite{chen2019,chen2021}.

}
    \end{abstract}

{Electrical control of information transport through interconnects in integrated circuits is an important goal for next-generation electronics that may be achieved with spintronic devices\upcite{liu2011spin,miron2011,liu2012spin,chen2018,chen2022,chen2023,mundy2017,baek2018complementary,baek2018spin,ryu2022,choi2022,kang2021,puliafito2021,rodrigues2023,yoon2023,fan2023}. Moving electrons may carry the spin information, but at the cost of Joule heating and relatively fast spin relaxation. Magnetic insulators enable pure spin transport via magnons\upcite{han2019,wang2019,wang2024}—the quanta of the spin wave excitations of the magnetic order—over micron-scale distances without moving charges, offering a promising route for low-power spintronics. In a widely used experimental setup\upcite{kajiwara2010, Cornelissen2015} the spin Hall effect\upcite{liu2011spin,liu2012spin,das2018} in a heavy metal (e.g., Pt) injects a magnonic spin current into an adjacent magnetic insulator (Figure~1a), which is subsequently detected as an electric voltage via the inverse spin Hall effect in a second Pt strip (Figure~1b). This framework extends to diverse insulating systems， including antiferromagnets\upcite{lebrun2018,han2020,huang2024}, paramagnets\upcite{oyanagi2019spin}, and even diamagnets\upcite{Schlitz2024}.

As {an alternative}, we propose ferroelectric materials %a possible solution
with spontaneous electric polarization that can be switched by electric fields. Despite being ubiquitous, e.g. as high-k insulators, they have never been considered as an information conduit in microelectronics. Recent theoretical studies\upcite{bauer2021,tang2022, tang2024} predict the existence of ferrons, quasiparticle excitations of the ferroelectric order that should enable electric polarization transport\upcite{bauer2021,tang2022a}, electric-field-tunable heat transport\upcite{tang2022,Brandi2023SA,bauer2023}, and nonlocal drag-induced thermoelectricity\upcite{tang2023}. However, to date the ferron theory remains untested due to the lack of direct experimental evidence.     %Ferroelectrics are characterized by spontaneous electric polarization that can be switched by an external electric field and, despite close analogies with magnetic materials, has so far received little attention in spintronics. % A ferroelectric (FE)-ferromagnetic (FM) bilayer, known as a composite multiferroic, offers an alternative to a single-phase multiferroic for multifunctional electronic devices due to its substantial emerging magnetoelectric coupling near interfaces. 
%Recent theoretical studies predicted the quasiparticle excitations of ``ferrons" in ferroelectric materials, quite analogous to magnons, which enable the electric polarization Seebeck effect, electric-field-tunable heat transport, and nonlocal thermoelectricity. Despite their theoretical potential, direct experimental evidence of ferrons has remained elusive, leaving their capabilities largely unexplored.}

Here we report %evidence of 
long-range transport of electric polarization in ferroelectric insulators at room temperature. %A configuration with two Pt contacts would give a null result since spins cannot be injected into non-magnetic insulators. Figure 1 schematizes the ferron injection-detection mechanism using ferromagnetic permalloy (Py) electrodes, leveraging dynamical magnetoelectric coupling via the spin Hall effect and spin-dependent conductivity. Instead, 
Figure 1 illustrates the mechanisms for ferron injection and detection. Ferromagnetic metal permalloy (Py) electrodes serve as ferron injectors and detectors, employing the dynamical magnetoelectric coupling induced by the combined %(anomalous) 
(anomalous) spin Hall effect and spin-dependent conductivity. %, as illustrated in Fig.~1. 
An applied electric current along \textit{y}-axis generates an anomalous Hall charge current and a transverse spin current along \(x\)-axis when the magnetization \textbf{m} is perpendicular to the plane (along $z$-axis). The former causes a steady state charge accumulation at the edges and dc electric stray fields that {cannot inject non-equilibrium ferrons or excite the detector contact}. The spin current, on the other hand, generates a non-equilibrium spin accumulation \(\mu_{s}\) polarized along \textbf{m} at the edges on the length scale of the spin-diffusion length with finite gradient  \( \partial _x \mu_{s}\). Spin filtering by metallic ferromagnet leads to a chemical potential gradient $-\alpha_{P}\partial _x \mu_{s}\mathbf{m}\times\mathbf{\hat{y}}$ (along $x$-axis when $\mathbf{m}\parallel
\mathbf{\hat{z}}$), where $\alpha_{P}$ %$=(\sigma_{\uparrow}-\sigma_{\downarrow})/(\sigma_{\uparrow}+\sigma_{\downarrow})$ 
is the spin polarization ratio of electric conductivities. The associated charge fluctuations are equivalent with a non-equilibrium electric field $\boldsymbol{\mathcal{E}}=-(\alpha_{P}/e)\partial _x \mu_{s}\mathbf{\hat{x}}$ that excites the ferrons in a ferroelectric with polarization \(\mathbf{P}\) via the dipolar Stark interaction $\mathcal{H}_{int}=-\boldsymbol{\mathcal{E}}\cdot\mathbf{P}$ (Figure~1{c}). This process is analogous to the excitation of magnons in a ferromagnet by the spin accumulation in a nonmagnetic contact mediated by the interface exchange interaction (Figure~1a). 

The injected ferrons diffuse from the injector to the detector, where the inverse process occurs. The ferron accumulation induces an effective non-equilibrium electric field $\mathcal{E}^{\prime}\mathbf{\hat{x}}$ in the detector magnet that generates an anomalous Hall current along the negative direction of $y$-axis (Figure~1d). %The detected current $j_{\textit{DET}}$ exhibits a direction opposite to that in the magnon-mediated transport scenario whereas the direction of the injected current $j_{\textit{INJ}}$ in Figure 1a and 1c maintains the same.
A striking distinction from magnon transport emerges in current-directionality: ferron detection/injection currents are antiparallel (Figure 1c, 1d), whereas magnon transport features parallel currents (Figure 1a, 1b). This polarity difference originates from the symmetry of dynamical magnetoelectric coupling: nonequilibrium ferron accumulation induces an effective electric field in the detector, collinear with the injector’s field orientation. This collinearity generates an anomalous Hall current in the detector opposite to the injector’s drive current, giving rise to the observed antiparallel polarity. % through which an effective transverse field (or a chemical potential gradient) is induced in the detector in the same direction as that in the injector by a nonequilibrium ferron accumulation, generating an anomalous Hall current opposite to the applied current in the injector. %Strikingly, in ferron transport scenaio, the directions of detecting and injecting currents are antiparallel (Figure 1c, 1d), while the counterparts are parallel for magnon transport (Figure 1a, 1b).} 
Crucially, ferron excitation requires the electric-field component parallel to the interface; rotating \textbf{m} from out-of-plane to in-plane eliminates this component, switching off the ferron current %Since only the component \(\mathcal{E}_x\) parallel to the interface can excite ferrons, the ferron current can be switched off by rotating the magnetization $m$ from out-of-plane to in-plane direction 
(Extended Data Figure 1). %or electrically by a 90 degree rotation of the polarization from the \(x\)-direction.    

    The experimental setup (Figure 2a) mirrors classic non-local magnon transport architectures\upcite{kajiwara2010, Cornelissen2015}. Two Py strips  {with an edge-to-edge distance $d$ = 1.5 $\mu$m} inject and detect ferrons in the ferroelectric lead magnesium niobate-lead titanate single crystal %film 
    {(PbMg$_{1/3}$Nb$_{2/3}$O$_3$)$_{0.7}$-(PbTiO$_3$)$_{0.3}$}
    (PMN-PT) %with %in-plane 
    %electric polarization along $x$-axis 
    (see Methods). %PSTO film. 
%ensures that any known effects does not influence the signals in the experiment except those about spin. 
    %The distance between two permalloy strips is $1.5{\mu}m$, which shows zero leakage current. 
    %Ferromagnetic permalloy is with high spin-orbit coupling for anomalous spin Hall effect (ASHE). Unlike the spin Hall effect (SHE), which only generates out-of-plane spin current, ASHE allows manipulating the direction of the generated spin current %%which is given by $\mathbf{M}{\times}\mathbf{I}$, 
    %through changing the direction of the external magnetic field. 
    %One of the permalloy strips acts as injector and the another one as detector. 
 %To achieve an output signal in the detector as large as possible, the input current in the injector is set as 2mA, which is already close to the breakdown current of the permalloy strips. 
    %Due to ASHE, a spin current $J_s$, which direction is given by the direction of $\mathbf{M}{\times}\mathbf{I}$, is generated in the injector and causes spin accumulation at both ends of it. The direction of the accumulated spin is perpendicular to $J_s$. 
    %direction of injected spin angular momentum. 
   A 3 T external magnetic field aligns the permalloy magnetization along the field direction, enabling control over equilibrium magnetization and nonequilibrium spin accumulation. %We control the equilibrium magnetization and thereby the non-equilibrium spin accumulation by a strong external magnetic field of 3 Tesla that forces the magnetization of the permalloy strips into the field direction.  %the maximum output signal. 
    %%When we enabled a charge current through the injector, a spin current is generated. We manipulate the direction of spin current in the injector and observe the first harmonic response 
    To isolate the ferron-mediated signal, we focus on the current-asymmetric nonlocal voltage $V_{nl}^{odd}=[V_{nl}(I)-V_{nl}(-I)]/2$, which eliminates thermal contributions from the anomalous Nernst effect (independent of 
I
 direction). %{Since a large contribution from the anomalous Nernst effect in the Py detector does not depend on the direction of the dc current \(I\) in the injector,}  we focus here on the transport signal $V_{nl}^{odd}=[V_{nl}(I)-V_{nl}(-I)]/2$ that is odd in the current direction, where $V_{nl}(I)$ is the nonlocal voltage. %In the coordinate system of Figure 1, to leading order in relevant magnetization component \(m_z\) and ferroelectric polarization component \(P_x\), 
    $V_{nl}^{odd}$ can be expressed as (see Methods)
\begin{equation}
V_{nl}^{odd}\approx A m_{z}^2 I +Bm_{z} P_{x} I^3
\end{equation}
  where %\(m_z\) is the relevant magnetization component, \(P_x\) is the component of ferroelectric polarization along $x$-axis, 
  $A$ and $B$ are constant coefficients, $m_z=m\cos\theta$ ($\theta$ is the 
 angle between \textbf{m} and \(z\)-axis (in the \(xz\)-plane)), $P_x$ is the ferroelectric polarization along $x$-axis. %we rotate the sample in external field to manipulate the direction of injected spin angular momentum. %%The angel $\theta$ between the direction of external field and the out-of-plane direction. 
  %As argued above, $V_{nl}^{odd}$ depends on the angle $\theta$ between \textbf{m} and \(z\)-axis (in the \(xz\)-plane) by the dynamical magnetoelectrical coupling at the PMN-PT $|$ Py interface and the ferron transport between the contacts. %To leading order in the current, 
  According to equation (1), the dependence on $m_z=m\cos\theta$ decomposes $V_{nl}^{odd}$ into two components, $V_{m_z^2}$ (even in $m_z$)
  and $V_{m_z}$ (odd in $m_z$). %=V_{m_z^2}+V_{m_z}$ has components that are even and odd in $m_z=M_s \cos{\theta}$. 
  {We obtain the largest $V_{nl}^{odd}$ for a current density of 8.9$\times$10$^6$ A cm$^{-2}$ above which  the injector becomes unstable.} 
  We fit the experimental raw data of {$V_{nl}$} (Extended Data Figure 2) with $a\cos^2\theta+b\cos\theta+c$ ($c$ is the offset), leading to the plots of {$V_{m_z^2}$ ($\cos^2\theta$ dependence) and $V_{m_z}$} ($\cos\theta$ dependence) in Figure 2b and 2c as a function of $\theta$.} 
  %As shown in Figure~\ref{fig:2}b, %%we observe the variation and demonstrate the spin transport in ferroelectrics.$\theta$-dependent $V_{nl}$ behaves like a superposition of trigonometric functions in different periods instead of a constant value, which is consistent with our former expectation. 
  %Both $V_{m_z^2}$ %\propto \cos{^2\theta}$ %(Figure~2d) 
  %is similar to the spin transport by magnons through magnetic insulators\upcite{Cornelissen2015}. Moreover, {
  %and $V_{m_z}$ %in Figure~2c changes sign when switching the out-of-plane component of magnetization. 
  %reach the peak under out-of-plane magnetic field and equal zero under in-plane magnetic field, which consists with the previous conclusion. The distinction lies in that $V_{m_z^2}$ keeps invariant in contrast to the reversed sign of $V_{m_z}$ under inverse out-of-plane magnetic field along $\hat{z}$ and $-\hat{z}$. 
  Moreover, Figure 2d and 2e illustrate the current density dependence of the difference between peak and valley of $V_{m_z^2}$ and $V_{m_z}$. $\Delta{V_{m_z^2}}=V_{m_z^2}(\theta=0^{\circ})-V_{m_z^2}(\theta=90^{\circ}) \propto j_{\textit{INJ}}$, 
while $\Delta{V_{m_z}}=V_{m_z}(\theta=0^{\circ})-V_{m_z}(\theta=180^{\circ})\propto j^3_{\textit{INJ}}$, consistent with ferron transport equation (1). 
    
 Direct evidence for ferron transport %external electric fields are applied to switch 
 arises from the the electric-field reversal of  ${V_{m_z}}$, enabled by initial voltage pulses to switch the ferroelectric polarization $P_x$. %The obtained data %after switching the in-plane polarization of PMN-PT along the $x$-axis by an initial large voltage pulse across the strips 
 %provides another strong evidence for the ferron hypothesis. 
  The PMN-PT is [001]-oriented  with [111] easy polarization axis. An {27 kV cm$^{-1}$} electric field can fully switch \(P_x\). %component. %by offering an external electric field over the coercivity of PMN-PT. 
    %On the basis of function (1), only the $V_{m_z}$ component is manipulatable by $P_x$. 
    %We further measure $\theta$ dependence of $V_{m_z}$ under inverse $P_x$. 
    In Figure~\ref{fig:3}a shows the $\theta$ dependence of $V_{m_z}$ under negative $P_x$, yielding $V_{m_z}({\theta=0^{\circ}})< 0$ and $V_{m_z}({\theta=180^{\circ}})>0$, %show different signs, %for {negative} \(P_x\) 
    which is consistent with the symmetry observed in  Figure 2b. %Besides, $V_{m_z}({\theta=0^{\circ}})<V_{m_z}(\theta=180^{\circ})$ is confirmed under negative $P_x$. 
    Polarization reversal flips the nonlocal signal: with positive $P_x$ (Figure 3b), $V_{m_z}({\theta=0^{\circ}})>0$ and $V_{m_z}({\theta=180^{\circ}})<0$ , demonstrating a strict correlation between $P_x$ and $V_{m_z}$ polarity. 
    %Furthermore, we vary the external out-of-plane magnetic field along $\hat{z}$ continuously from +3 Tesla to -3 Tesla and backwards. $V_{nl}$ are split into an even function $V_{m_z^2}$ and an odd function $V_{m_z}$ in $m_z$ (Extended Data Figure 2). Then we observe $V_{m_z}$ as a function of perpendicular magnetic field. The results shown in Figure~\ref{fig:3}c and 3d reconfirm the electric field switching of $V_{m_z}$. 
    %$V_{m_z}$ as a function of perpendicular magnetic field in Figure 3c and 3d, reconfirm the electric-field switching of $V_{m_z}$.
    Furthermore, Figure 3c and 3d show $V_{m_z}$ as a function of perpendicular magnetic field. The output signal under saturated out-of-plane magnetizations switches sign when we reverse $P_x$, confirming the electric field switching of $\Delta{V_{m_z}}$.
    In Figure 3e, we plot $\Delta{V_{m_z}}$ %$ = V_{m_z}(\theta=0^{\circ})-V_{m_z}(\theta=180^{\circ})$ 
    at zero electric field after repeatedly applying %representative electric fields 
    {-27 and +27 kV cm$^{-1}$} to PMN-PT over four cycles. Indeed, $\Delta{V_{m_z}}$  
    is negative after {-27} kV cm$^{-1}$, and positive after {+27} kV cm$^{-1}$. In contrast, $\Delta{V_{m_z^2}}$ does not show \(P_x\) dependence (Extended Data Figure 3). 
    %Figure 3d summarizes the results for $\Delta{V_{m_z}}$ 
    %when \(E\) is cycled from {+27 to -27 and then back to +27} kV cm$^{-1}$. 
    The most prominent feature is the hysteresis loop shown in Figure 3f, % representing electric field dependence of $\Delta{V_{m_z}}$, in which the threshold of %switching 
    %electric field matches the coercive field of PMN-PT.
    where $\Delta{V_{m_z}}$ tracks the applied electric field with a threshold matching the coercive field of PMN-PT—an unambiguous fingerprint of ferron-mediated transport coupled to ferroelectric polarization reversal.

Figure 4 presents $\Delta{V_{m_z}}$ as a function of distance $d$ between Py injector and detector, which follows ferron diffusion model %equation 
(see Methods) 
\begin{equation}
    	\Delta{V_{{m_z}}} = C \mathrm{e} ^{-d/\lambda},
\end{equation}
    where $C$ is a constant, yielding a ferron diffusion length $\lambda\approx{14.8\pm1.2 \,\mathrm{\mu m}}$. The dependence of $\Delta{V_{m_z^2}}$ also matches well with this model (Extended Data Figure 4). %, but the calculated diffusion length $\lambda\approx{4.0\pm0.3\,\mathrm{\mu m}}$, different from the former result. %Since other factors, e.g. phonons, may also %be the media of spin transport cause non-local voltage response in devices composed of ferromagnetic metal injector and detector strips on non-magnetic insulator\upcite{Schlitz2024}, their contributions in $V_{m_z^2}$ cannot be certainly excluded, which may account for the difference of $\lambda$ of $V_{m_z}$ and $V_{m_z^2}$. 
 The mechanism we propose here differs from the recently predicted dynamical multiferroicity, which envisions a magnetization induced by a time-dependent electric polarization in non-ferromagnetic materials{\upcite{juraschek2017,basini2024}}. {Although the phenomenology of our signals is similar to the non-local transmission between two ferromagnetic contacts through a dielectric and diamagnetic insulator\upcite{Schlitz2024}, ours is orders of magnitude larger for comparable configurations, highlighting the unique efficiency of ferron-mediated transport in ferroelectrics..} %channel. 

    In summary, we demonstrate that "ferrons"—quasiparticle excitations of electric dipolar order in ferroelectrics—mediate long-range transport at room temperature via dynamical magnetoelectric coupling across ferroelectric-ferromagnetic interfaces. By interfacing ferroelectric PMN-PT with Py spin injectors and detectors, we show that nonequilibrium spin currents in Py generate effective electric fields through spin accumulation, exciting ferrons that diffuse over micrometer distances and induce detectable signals in remote detectors. The transport efficiency is reversibly controlled by both electric fields (via polarization switching) and magnetic fields (via contact magnetization orientation), with signal polarity strictly tied to the ferroelectric polarization direction. These findings establish ferrons as viable carriers for low-dissipation information transport, paving the way for "ferronics"—a new paradigm of ferroelectric-based devices integrating electric-field and temperature-gradient control for multifunctional, magnetoelectrically tunable applications.%Our findings redefine information transport paradigms in non-magnetic materials and establish ferroelectrics as a versatile platform for energy-efficient information transmission.%In summary, spin transport within ferroelectric materials is verified through the identification of ferrons—quasiparticles representing the collective excitation of electric dipole order.  Unlike traditional spin transport in ferromagnetic metals, which relies on spin-polarized current and incurs power loss due to Joule heating, ferron-mediated spin transport in ferroelectric insulators enables low-power operation with giant efficiency and um-range diffusion length. Ferrons, akin to ferroelectric polarized phonons, can be circularly polarized to transmit spin angular momentum and are fully manipulable by external electric fields, allowing for voltage-controlled device operation. These results solidify the existence of ferrons as spin transport carriers and pave the way for ferron-based spintronic devices, offering an alternative approach to surpass the limitations of CMOS technology and traditional spintronics.
    
    \newpage
    \begin{figure}[t!]
    	\centering
    	\includegraphics[width=0.8\textwidth]{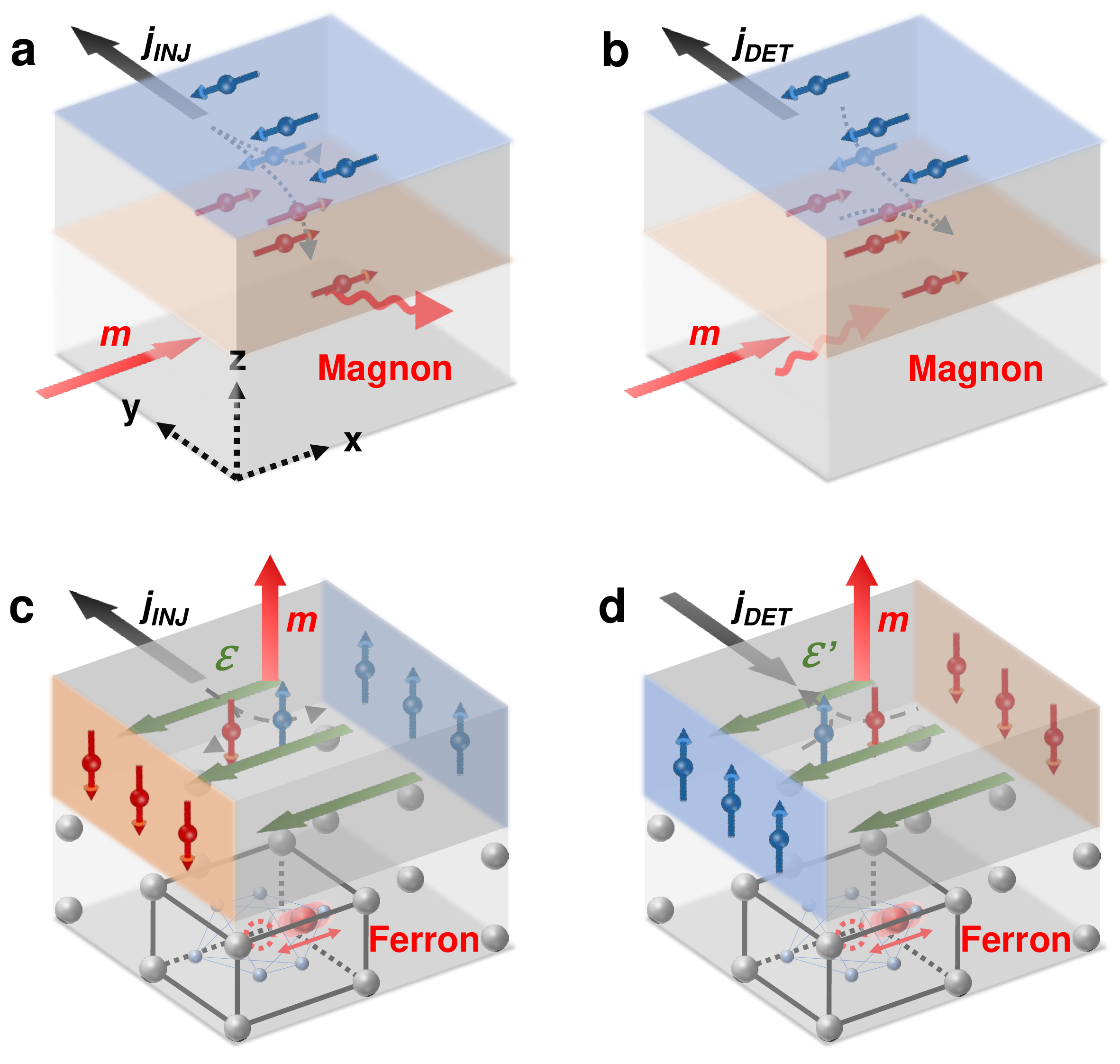}
    \caption{\textbf{Ferron excitation and detection. a}, \textbf{b}, The magnon excitation and detection in magnetic insulators with in-plane magnetization %in the plane 
    by the spin Hall effect and inverse spin Hall effect in a heavy metal contacts. The injected current density $j_{\textit{INJ}}$ and the detected current density $j_{\textit{DET}}$ are parallel along $y$-axis. %compared with
    \textbf{c}, The ferron excitation by a ferromagnetic metal contact on top of a ferroelectric film, where \(j_\textit{INJ}\) is a charge current bias. An out-of-plane magnetization %out of the plane 
    leads to an inhomogeneous spin accumulation ($\mu_{s}$) by the spin Hall effect at the edges. The spin-charge coupling of conduction electron currents in ferromagnets causes a fluctuating non-equilibrium electric field $\mathcal{E}\propto\alpha_{P}\nabla\mu_{s}$ where $\alpha_{P}=(\sigma_{\uparrow}-\sigma_{\downarrow})/(\sigma_{\uparrow}+\sigma_{\downarrow})$ and $\sigma_{\uparrow}$ $(\sigma_{\downarrow})$ are the electric conductivities for spin-up and -down electrons, that excite ferrons in the ferroelectric. \textbf{d}, The ferron accumulation diffuses towards the detector ferromagnetic strip, and a reciprocal process induces an electric current \(j_\textit{DET}\) along the negative direction of $y$-axis %in the opposite direction 
    antiparallel to $j_{\textit{INJ}}$ along $y$-axis, whereas the detected and injected currents are parallel in the magnon-mediated transport scenario depicted in panels \textbf{a} and \textbf{b}.} 
    	\label{fig:1}
    	%\vspace{80pt}
    \end{figure}

    \begin{figure}[t!]
    	\centering
    	\includegraphics[width=0.8\textwidth]{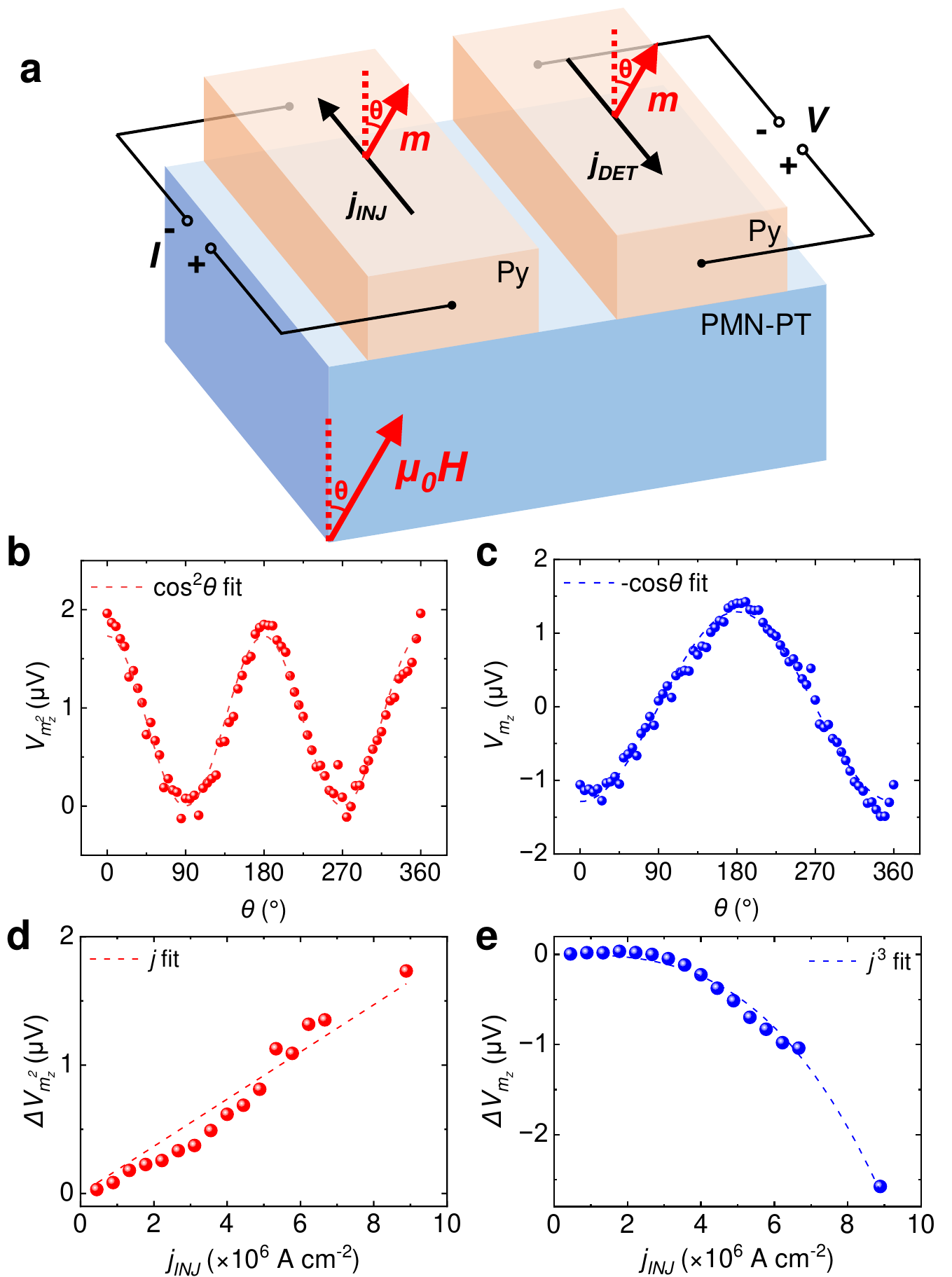}
    	\caption{\textbf{%Spin transport in ferroelectrics.
        Symmetry-resolved {nonlocal signal} %harmonic signals 
        of ferron transport. a,} Schematic representation of the sample configuration of the non-local measurements (not to scale). Two Py strips deposited onto a PMN-PT ferroelectric substrate serve as spin injector and detector. An external magnetic field $H$ controls the direction of the magnetization $\mathbf{m}$ and angle \(\theta\). %Two parallel permalloy strips are plated on the ferroelectric PMN-PT. A charge current flows through the left permalloy strip and causes in-plane spin accumulation in it via anomalous spin Hall effect (ASHE). This kind of spin accumulation leads to diffusion of angular momentum from permalloy to PSTO. Hence the concentration gradient of circularly polarized ferrons in PSTO is created. Through diffusion of ferrons, angular momentum is transmitted again from PSTO to the right permalloy strip which is 1.5$\mu$m away from the left one and gives rise to spin accumulation. By way of the converse effect of ASHE, an output current is generated and could be detected by the variation of the output voltage. 
        \textbf{b,} Angle dependence of {$V_{m_z^2}$ %of the response
        }to the current in the injector %\((\propto I)\) 
        of the non-local voltage (dots). The red curve is a $\cos{^2\theta}$ fit. \textbf{c,} Angle dependence of the non-local voltage signal $V_{m_z}$%\propto I^3$ }  
        fitted by $\cos{\theta}$, which changes sign upon switching the out-of-plane magnetization. \textbf{d,} $V_{m_z^2}$ linearly depends on the applied current density $j_\textit{INJ}$. \textbf{e,} Cubic current dependence of $V_{m_z}$, confirming its nonlinearity. }%Comparison between magnetic field dependence of the first harmonic response of non-local voltage signal under in-plane field and out-of-plane field. \textbf{d,} Angle dependence of ${\Delta}V_{out}$. ${\Delta}V_{out}$ represents the gap between $V_{nl}$ under zero field and saturation field.} 
    	\label{fig:2}
    	%\vspace{80pt}
    \end{figure}

    \begin{figure}[t!]
    	\centering
    	\includegraphics[width=1\textwidth]{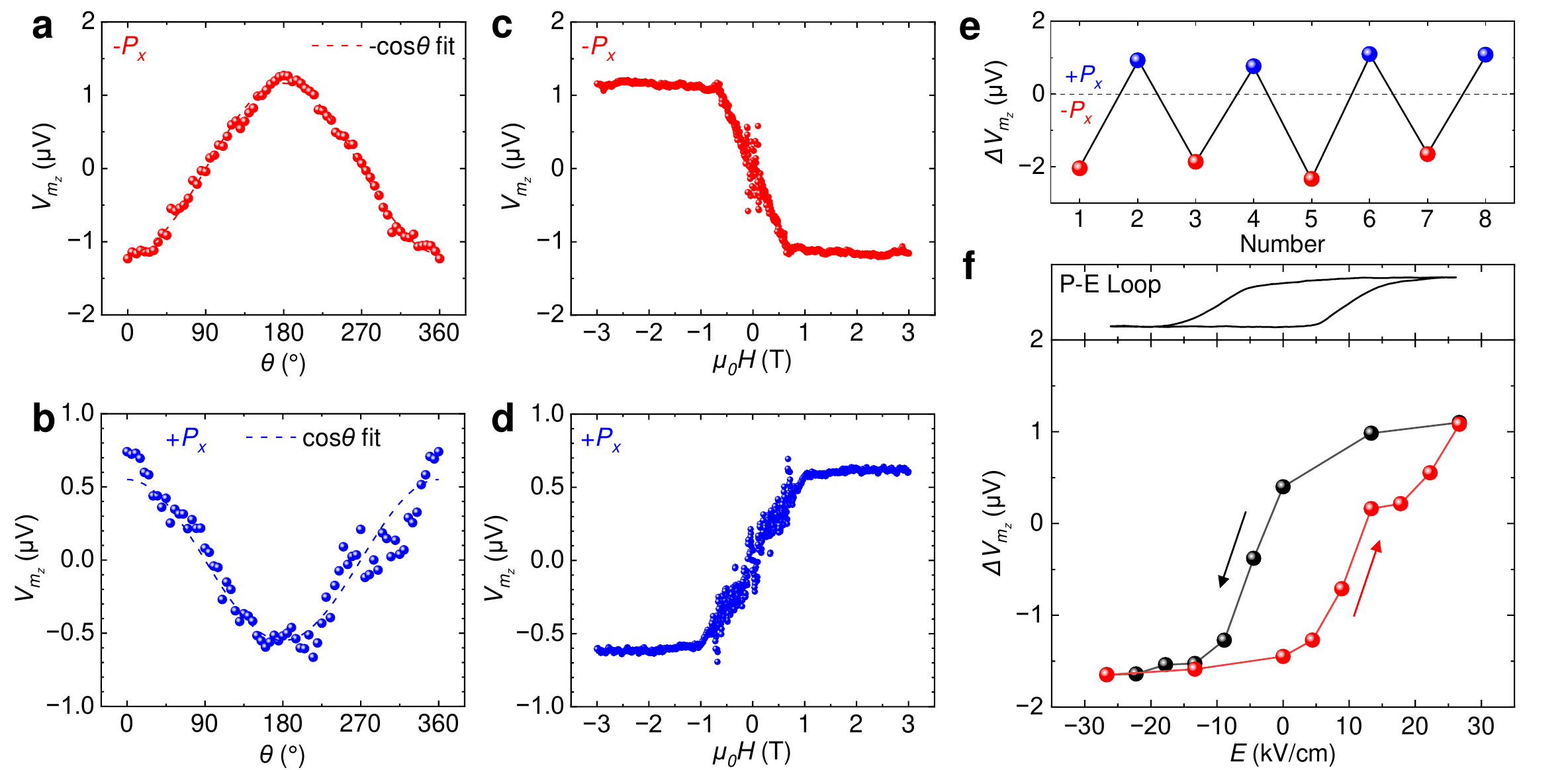}
    	\caption{\textbf{Electric field control of ferron transport. %Spin transport modulation in ferroelectrics by external electric field. 
        a}, \textbf{b}, $V_{m_z}$ as a function of magnetization angle $\theta$ for opposite ferroelectric polarizations, demonstrating polarization-dependent sign reversal. %Schematic representation. The external voltage is added to polarize the local ferroelectric film. Then the non-local signal is measured. The electric field is considered as a uniform and calculated by the added external voltage. 
        \textbf{c}, \textbf{d}, The out-of-plane magnetic field-dependent {$V_{m_z}$} under opposite ferroelectric polarization is consistent with the polarity shown in \textbf{a}, \textbf{b}. \textbf{e,} The $V_{m_z}$ modulation is robust over four electric field cycles.
        \textbf{f,} Upper panel shows electric field dependence of the polarization of PMN-PT, while lower panel illustrates the electric field dependence of $V_{m_z}$. $V_{m_z}$ as a function of electric field is a hysteresis loop.} %Electric field dependence of the polarization of PSTO.%the gap between $V_{out}$ under maximum magnetic field in reversed directions.  %Magnetic field dependence of $V_{nl}$ when the external electric field equals $3kV/cm$ and $-3kV/cm$.
    	\label{fig:3}
    	%\vspace{80pt}
    \end{figure}

    \begin{figure}[t!]
    	\centering
    	\includegraphics[width=1\textwidth]{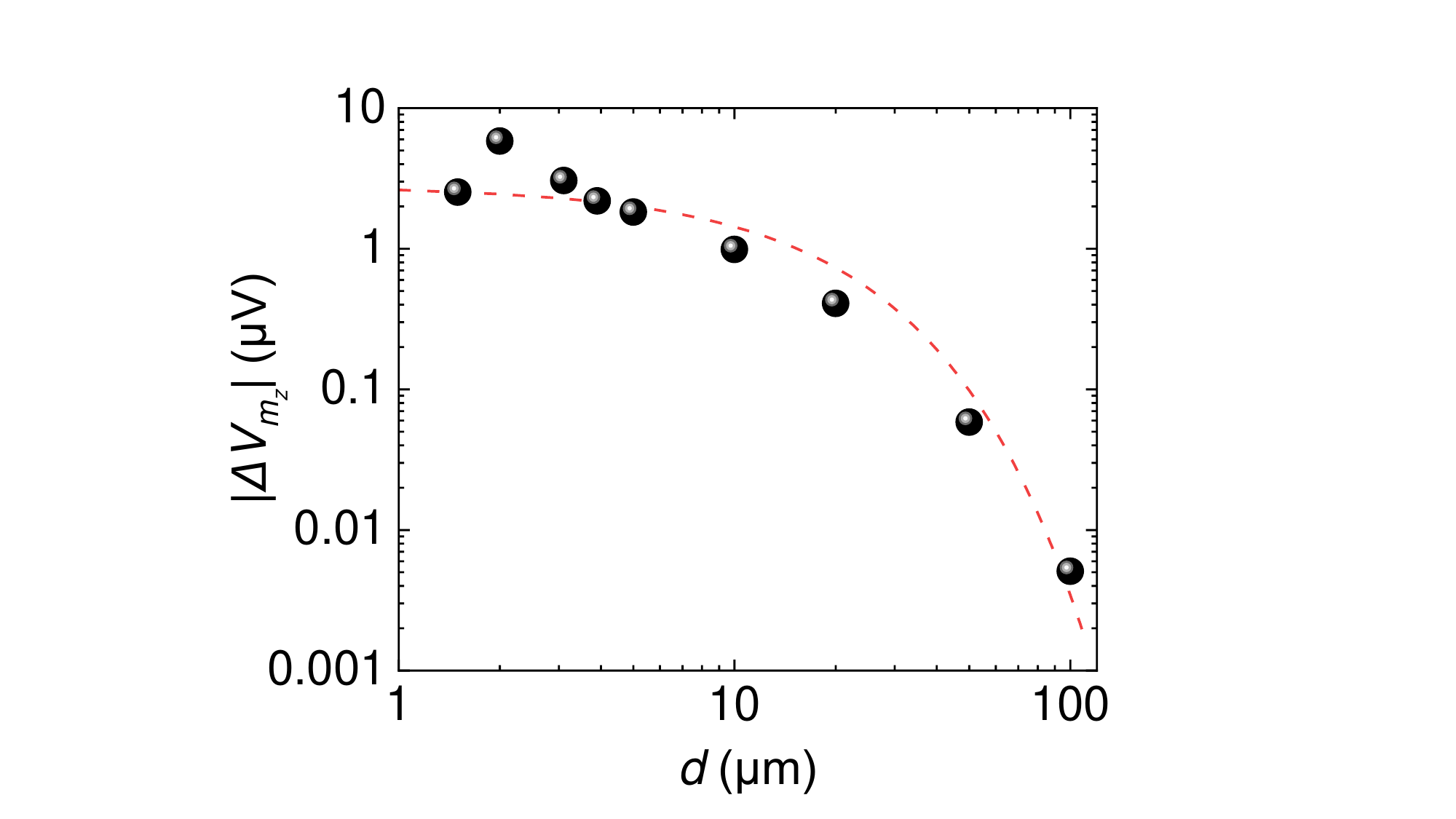}
    	\caption{\textbf{Long-range diffusive transport of ferrons. }Dependence of the nonlocal signal $V_{m_z}$ on the injector-detector spacing $d$. Data points (dots) fit the diffusive transport model (red curve), yielding a ferron diffusion length of $14.8\pm1.2$ $\mu$m.} %Schematic representation. The external voltage is added to polarize the local ferroelectric film. Then the non-local signal is measured. The electric field is considered as a uniform and calculated by the added external voltage. \textbf{b,} Electric field dependence of the polarization of PSTO. \textbf{c,} Magnetic field dependence of $V_{nl}$ when the external electric field equals $3kV/cm$ and $-3kV/cm$. \textbf{d,} Electric field dependence of ${\Delta}V$+-, which is the gap between $V_{out}$ under maximum magnetic field in reversed directions. } 
    	\label{fig:4}
    	%\vspace{80pt}
    \end{figure}

    \FloatBarrier

    \noindent\textbf{References}
    \bibliographystyle{naturemag}
    \bibliography{bibliography}

\begin{thebibliography}{10}
\expandafter\ifx\csname url\endcsname\relax
  \def\url#1{\texttt{#1}}\fi
\expandafter\ifx\csname urlprefix\endcsname\relax\def\urlprefix{URL }\fi
\providecommand{\bibinfo}[2]{#2}
\providecommand{\eprint}[2][]{\url{#2}}

\bibitem{wei2018}
\bibinfo{author}{Wei, Y.} \emph{et~al.}
\newblock \bibinfo{title}{A rhombohedral ferroelectric phase in epitaxially strained {H}f$_{0.5}${Z}r$_{0.5}${O}$_{2}$ thin films}.
\newblock \emph{\bibinfo{journal}{Nat. Mater.}} \textbf{\bibinfo{volume}{17}}, \bibinfo{pages}{1095--1100} (\bibinfo{year}{2018}).

\bibitem{wang2023}
\bibinfo{author}{Wang, Y.} \emph{et~al.}
\newblock \bibinfo{title}{A stable rhombohedral phase in ferroelectric {H}f({Z}r)$_{1+x}${O}$_{2}$ capacitor with ultralow coercive field}.
\newblock \emph{\bibinfo{journal}{Science}} \textbf{\bibinfo{volume}{381}}, \bibinfo{pages}{558--563} (\bibinfo{year}{2023}).

\bibitem{meier2012}
\bibinfo{author}{Meier, D.} \emph{et~al.}
\newblock \bibinfo{title}{Anisotropic conductance at improper ferroelectric domain walls}.
\newblock \emph{\bibinfo{journal}{Nat. Mater.}} \textbf{\bibinfo{volume}{11}}, \bibinfo{pages}{284--288} (\bibinfo{year}{2012}).

\bibitem{tang2022}
\bibinfo{author}{Tang, P.}, \bibinfo{author}{Iguchi, R.}, \bibinfo{author}{Uchida, K.-i.} \& \bibinfo{author}{Bauer, G.~E.}
\newblock \bibinfo{title}{Excitations of the ferroelectric order}.
\newblock \emph{\bibinfo{journal}{Phys. Rev. B}} \textbf{\bibinfo{volume}{106}}, \bibinfo{pages}{L081105} (\bibinfo{year}{2022}).

\bibitem{tang2022a}
\bibinfo{author}{Tang, P.}, \bibinfo{author}{Iguchi, R.}, \bibinfo{author}{Uchida, K.-i.} \& \bibinfo{author}{Bauer, G. E.~W.}
\newblock \bibinfo{title}{Thermoelectric polarization transport in ferroelectric ballistic point contacts}.
\newblock \emph{\bibinfo{journal}{Phys. Rev. Lett.}} \textbf{\bibinfo{volume}{128}}, \bibinfo{pages}{047601} (\bibinfo{year}{2022}).

\bibitem{tang2023}
\bibinfo{author}{Tang, P.}, \bibinfo{author}{Uchida, K.-i.} \& \bibinfo{author}{Bauer, G. E.~W.}
\newblock \bibinfo{title}{Nonlocal drag thermoelectricity generated by ferroelectric van der waals heterostructures}.
\newblock \emph{\bibinfo{journal}{Phys. Rev. B}} \textbf{\bibinfo{volume}{107}}, \bibinfo{pages}{L121406} (\bibinfo{year}{2023}).

\bibitem{tang2024}
\bibinfo{author}{Tang, P.} \& \bibinfo{author}{Bauer, G.~E.}
\newblock \bibinfo{title}{Electric analog of magnons in order-disorder ferroelectrics}.
\newblock \emph{\bibinfo{journal}{Phys. Rev. B}} \textbf{\bibinfo{volume}{109}}, \bibinfo{pages}{L060301} (\bibinfo{year}{2024}).

\bibitem{bauer2021}
\bibinfo{author}{Bauer, G. E.~W.}, \bibinfo{author}{Iguchi, R.} \& \bibinfo{author}{Uchida, K.-i.}
\newblock \bibinfo{title}{Theory of transport in ferroelectric capacitors}.
\newblock \emph{\bibinfo{journal}{Phys. Rev. Lett.}} \textbf{\bibinfo{volume}{126}}, \bibinfo{pages}{187603} (\bibinfo{year}{2021}).

\bibitem{bauer2022}
\bibinfo{author}{Bauer, G.~E.}, \bibinfo{author}{Tang, P.}, \bibinfo{author}{Iguchi, R.} \& \bibinfo{author}{Uchida, K.-i.}
\newblock \bibinfo{title}{Magnonics vs. ferronics}.
\newblock \emph{\bibinfo{journal}{J. Magn. Magn. Mater.}} \textbf{\bibinfo{volume}{541}}, \bibinfo{pages}{168468} (\bibinfo{year}{2022}).

\bibitem{bauer2023}
\bibinfo{author}{Bauer, G.~E.} \emph{et~al.}
\newblock \bibinfo{title}{Polarization transport in ferroelectrics}.
\newblock \emph{\bibinfo{journal}{Phys. Rev. Appl.}} \textbf{\bibinfo{volume}{20}}, \bibinfo{pages}{050501} (\bibinfo{year}{2023}).

\bibitem{chen2019}
\bibinfo{author}{Chen, X.} \emph{et~al.}
\newblock \bibinfo{title}{Electric field control of n{\'e}el spin--orbit torque in an antiferromagnet}.
\newblock \emph{\bibinfo{journal}{Nat. Mater.}} \textbf{\bibinfo{volume}{18}}, \bibinfo{pages}{931--935} (\bibinfo{year}{2019}).

\bibitem{chen2021}
\bibinfo{author}{Chen, X.} \emph{et~al.}
\newblock \bibinfo{title}{Observation of the antiferromagnetic spin hall effect}.
\newblock \emph{\bibinfo{journal}{Nat. Mater.}} \textbf{\bibinfo{volume}{20}}, \bibinfo{pages}{800--804} (\bibinfo{year}{2021}).

\bibitem{liu2011spin}
\bibinfo{author}{Liu, L.}, \bibinfo{author}{Moriyama, T.}, \bibinfo{author}{Ralph, D.} \& \bibinfo{author}{Buhrman, R.}
\newblock \bibinfo{title}{Spin-torque ferromagnetic resonance induced by the spin hall effect}.
\newblock \emph{\bibinfo{journal}{Phys. Rev. Lett.}} \textbf{\bibinfo{volume}{106}}, \bibinfo{pages}{036601} (\bibinfo{year}{2011}).

\bibitem{miron2011}
\bibinfo{author}{Miron, I.~M.} \emph{et~al.}
\newblock \bibinfo{title}{Perpendicular switching of a single ferromagnetic layer induced by in-plane current injection}.
\newblock \emph{\bibinfo{journal}{Nature}} \textbf{\bibinfo{volume}{476}}, \bibinfo{pages}{189--193} (\bibinfo{year}{2011}).

\bibitem{liu2012spin}
\bibinfo{author}{Liu, L.} \emph{et~al.}
\newblock \bibinfo{title}{Spin-torque switching with the giant spin hall effect of tantalum}.
\newblock \emph{\bibinfo{journal}{Science}} \textbf{\bibinfo{volume}{336}}, \bibinfo{pages}{555--558} (\bibinfo{year}{2012}).

\bibitem{chen2018}
\bibinfo{author}{Chen, X.} \emph{et~al.}
\newblock \bibinfo{title}{Antidamping-torque-induced switching in biaxial antiferromagnetic insulators}.
\newblock \emph{\bibinfo{journal}{Phys. Rev. Lett.}} \textbf{\bibinfo{volume}{120}}, \bibinfo{pages}{207204} (\bibinfo{year}{2018}).

\bibitem{chen2022}
\bibinfo{author}{Chen, X.} \emph{et~al.}
\newblock \bibinfo{title}{Control of spin current and antiferromagnetic moments via topological surface state}.
\newblock \emph{\bibinfo{journal}{Nat. Electron.}} \textbf{\bibinfo{volume}{5}}, \bibinfo{pages}{574--578} (\bibinfo{year}{2022}).

\bibitem{chen2023}
\bibinfo{author}{Chen, X.} \emph{et~al.}
\newblock \bibinfo{title}{Octupole-driven magnetoresistance in an antiferromagnetic tunnel junction}.
\newblock \emph{\bibinfo{journal}{Nature}} \textbf{\bibinfo{volume}{613}}, \bibinfo{pages}{490--495} (\bibinfo{year}{2023}).

\bibitem{mundy2017}
\bibinfo{author}{Mundy, J.~A.} \emph{et~al.}
\newblock \bibinfo{title}{Functional electronic inversion layers at ferroelectric domain walls}.
\newblock \emph{\bibinfo{journal}{Nat. Mater.}} \textbf{\bibinfo{volume}{16}}, \bibinfo{pages}{622--627} (\bibinfo{year}{2017}).

\bibitem{baek2018complementary}
\bibinfo{author}{Baek, S.-h.~C.} \emph{et~al.}
\newblock \bibinfo{title}{Complementary logic operation based on electric-field controlled spin--orbit torques}.
\newblock \emph{\bibinfo{journal}{Nat. Electron.}} \textbf{\bibinfo{volume}{1}}, \bibinfo{pages}{398--403} (\bibinfo{year}{2018}).

\bibitem{baek2018spin}
\bibinfo{author}{Baek, S.-h.~C.} \emph{et~al.}
\newblock \bibinfo{title}{Spin currents and spin--orbit torques in ferromagnetic trilayers}.
\newblock \emph{\bibinfo{journal}{Nat. Mater.}} \textbf{\bibinfo{volume}{17}}, \bibinfo{pages}{509--513} (\bibinfo{year}{2018}).

\bibitem{ryu2022}
\bibinfo{author}{Ryu, J.} \emph{et~al.}
\newblock \bibinfo{title}{Efficient spin--orbit torque in magnetic trilayers using all three polarizations of a spin current}.
\newblock \emph{\bibinfo{journal}{Nat. Electron.}} \textbf{\bibinfo{volume}{5}}, \bibinfo{pages}{217--223} (\bibinfo{year}{2022}).

\bibitem{choi2022}
\bibinfo{author}{Choi, J.-G.} \emph{et~al.}
\newblock \bibinfo{title}{Voltage-driven gigahertz frequency tuning of spin hall nano-oscillators}.
\newblock \emph{\bibinfo{journal}{Nat. Commun.}} \textbf{\bibinfo{volume}{13}}, \bibinfo{pages}{3783} (\bibinfo{year}{2022}).

\bibitem{kang2021}
\bibinfo{author}{Kang, J.} \emph{et~al.}
\newblock \bibinfo{title}{Current-induced manipulation of exchange bias in {I}r{M}n/{N}i{F}e bilayer structures}.
\newblock \emph{\bibinfo{journal}{Nat. Commun.}} \textbf{\bibinfo{volume}{12}}, \bibinfo{pages}{6420} (\bibinfo{year}{2021}).

\bibitem{puliafito2021}
\bibinfo{author}{Puliafito, V.}, \bibinfo{author}{Sanchez-Tejerina, L.}, \bibinfo{author}{Carpentieri, M.}, \bibinfo{author}{Azzerboni, B.} \& \bibinfo{author}{Finocchio, G.}
\newblock \bibinfo{title}{Modulation, injection locking, and pulling in an antiferromagnetic spin-orbit torque oscillator}.
\newblock \emph{\bibinfo{journal}{IEEE Trans. Magn.}} \textbf{\bibinfo{volume}{57}}, \bibinfo{pages}{1--6} (\bibinfo{year}{2021}).

\bibitem{rodrigues2023}
\bibinfo{author}{Rodrigues, D.~R.} \emph{et~al.}
\newblock \bibinfo{title}{Dynamical neural network based on spin transfer nano-oscillators}.
\newblock \emph{\bibinfo{journal}{IEEE Trans. Nanotechnol.}} \textbf{\bibinfo{volume}{22}}, \bibinfo{pages}{800--805} (\bibinfo{year}{2023}).

\bibitem{yoon2023}
\bibinfo{author}{Yoon, J.-Y.} \emph{et~al.}
\newblock \bibinfo{title}{Handedness anomaly in a non-collinear antiferromagnet under spin--orbit torque}.
\newblock \emph{\bibinfo{journal}{Nat. Mater.}} \textbf{\bibinfo{volume}{22}}, \bibinfo{pages}{1106--1113} (\bibinfo{year}{2023}).

\bibitem{fan2023}
\bibinfo{author}{Fan, Y.} \emph{et~al.}
\newblock \bibinfo{title}{Coherent magnon-induced domain-wall motion in a magnetic insulator channel}.
\newblock \emph{\bibinfo{journal}{Nat. Nanotechnol.}} \textbf{\bibinfo{volume}{18}}, \bibinfo{pages}{1000--1004} (\bibinfo{year}{2023}).

\bibitem{han2019}
\bibinfo{author}{Han, J.}, \bibinfo{author}{Zhang, P.}, \bibinfo{author}{Hou, J.~T.}, \bibinfo{author}{Siddiqui, S.~A.} \& \bibinfo{author}{Liu, L.}
\newblock \bibinfo{title}{Mutual control of coherent spin waves and magnetic domain walls in a magnonic device}.
\newblock \emph{\bibinfo{journal}{Science}} \textbf{\bibinfo{volume}{366}}, \bibinfo{pages}{1121--1125} (\bibinfo{year}{2019}).

\bibitem{wang2019}
\bibinfo{author}{Wang, Y.} \emph{et~al.}
\newblock \bibinfo{title}{Magnetization switching by magnon-mediated spin torque through an antiferromagnetic insulator}.
\newblock \emph{\bibinfo{journal}{Science}} \textbf{\bibinfo{volume}{366}}, \bibinfo{pages}{1125--1128} (\bibinfo{year}{2019}).

\bibitem{wang2024}
\bibinfo{author}{Wang, F.} \emph{et~al.}
\newblock \bibinfo{title}{Deterministic switching of perpendicular magnetization by out-of-plane anti-damping magnon torques}.
\newblock \emph{\bibinfo{journal}{Nat. Nanotechnol.}} \textbf{\bibinfo{volume}{19}}, \bibinfo{pages}{1478--1484} (\bibinfo{year}{2024}).

\bibitem{kajiwara2010}
\bibinfo{author}{Kajiwara, Y.} \emph{et~al.}
\newblock \bibinfo{title}{Transmission of electrical signals by spin-wave interconversion in a magnetic insulator}.
\newblock \emph{\bibinfo{journal}{Nature}} \textbf{\bibinfo{volume}{464}}, \bibinfo{pages}{262--266} (\bibinfo{year}{2010}).

\bibitem{Cornelissen2015}
\bibinfo{author}{Cornelissen, L.}, \bibinfo{author}{Liu, J.}, \bibinfo{author}{Duine, R.}, \bibinfo{author}{Youssef, J.} \& \bibinfo{author}{van Wees, B.}
\newblock \bibinfo{title}{{Long-distance transport of magnon spin information in a magnetic insulator at room temperature}}.
\newblock \emph{\bibinfo{journal}{Nat. Phys.}} \textbf{\bibinfo{volume}{11}}, \bibinfo{pages}{1022--1026} (\bibinfo{year}{2015}).

\bibitem{das2018}
\bibinfo{author}{Das, K.~S.}, \bibinfo{author}{Liu, J.}, \bibinfo{author}{van Wees, B.~J.} \& \bibinfo{author}{Vera-Marun, I.~J.}
\newblock \bibinfo{title}{Efficient injection and detection of out-of-plane spins via the anomalous spin hall effect in permalloy nanowires}.
\newblock \emph{\bibinfo{journal}{Nano Lett.}} \textbf{\bibinfo{volume}{18}}, \bibinfo{pages}{5633--5639} (\bibinfo{year}{2018}).

\bibitem{lebrun2018}
\bibinfo{author}{Lebrun, R.} \emph{et~al.}
\newblock \bibinfo{title}{Tunable long-distance spin transport in a crystalline antiferromagnetic iron oxide}.
\newblock \emph{\bibinfo{journal}{Nature}} \textbf{\bibinfo{volume}{561}}, \bibinfo{pages}{222--225} (\bibinfo{year}{2018}).

\bibitem{han2020}
\bibinfo{author}{Han, J.} \emph{et~al.}
\newblock \bibinfo{title}{Birefringence-like spin transport via linearly polarized antiferromagnetic magnons}.
\newblock \emph{\bibinfo{journal}{Nat. Nanotechnol.}} \textbf{\bibinfo{volume}{15}}, \bibinfo{pages}{563--568} (\bibinfo{year}{2020}).

\bibitem{huang2024}
\bibinfo{author}{Huang, X.} \emph{et~al.}
\newblock \bibinfo{title}{Manipulating chiral spin transport with ferroelectric polarization}.
\newblock \emph{\bibinfo{journal}{Nat. Mater.}} \bibinfo{pages}{1--7} (\bibinfo{year}{2024}).

\bibitem{oyanagi2019spin}
\bibinfo{author}{Oyanagi, K.} \emph{et~al.}
\newblock \bibinfo{title}{Spin transport in insulators without exchange stiffness}.
\newblock \emph{\bibinfo{journal}{Nat. Commun.}} \textbf{\bibinfo{volume}{10}}, \bibinfo{pages}{4740} (\bibinfo{year}{2019}).

\bibitem{Schlitz2024}
\bibinfo{author}{Schlitz, R.} \emph{et~al.}
\newblock \bibinfo{title}{{Electrically induced angular momentum flow between separated ferromagnets}}.
\newblock \emph{\bibinfo{journal}{Phys. Rev. Lett.}} \textbf{\bibinfo{volume}{132}}, \bibinfo{pages}{256701} (\bibinfo{year}{2024}).

\bibitem{Brandi2023SA}
\bibinfo{author}{Wooten, B.~L.} \emph{et~al.}
\newblock \bibinfo{title}{Electric field–dependent phonon spectrum and heat conduction in ferroelectrics}.
\newblock \emph{\bibinfo{journal}{Sci. Adv.}} \textbf{\bibinfo{volume}{9}}, \bibinfo{pages}{eadd7194} (\bibinfo{year}{2023}).

\bibitem{juraschek2017}
\bibinfo{author}{Juraschek, D.~M.}, \bibinfo{author}{Fechner, M.}, \bibinfo{author}{Balatsky, A.~V.} \& \bibinfo{author}{Spaldin, N.~A.}
\newblock \bibinfo{title}{Dynamical multiferroicity}.
\newblock \emph{\bibinfo{journal}{Phys. Rev. Mater.}} \textbf{\bibinfo{volume}{1}}, \bibinfo{pages}{014401} (\bibinfo{year}{2017}).

\bibitem{basini2024}
\bibinfo{author}{Basini, M.} \emph{et~al.}
\newblock \bibinfo{title}{Terahertz electric-field-driven dynamical multiferroicity in {S}r{T}i{O}$_{3}$}.
\newblock \emph{\bibinfo{journal}{Nature}} \textbf{\bibinfo{volume}{628}}, \bibinfo{pages}{534--539} (\bibinfo{year}{2024}).

\bibitem{avci2015}
\bibinfo{author}{Avci, C.~O.} \emph{et~al.}
\newblock \bibinfo{title}{Unidirectional spin hall magnetoresistance in ferromagnet/normal metal bilayers}.
\newblock \emph{\bibinfo{journal}{Nat. Phys.}} \textbf{\bibinfo{volume}{11}}, \bibinfo{pages}{570--575} (\bibinfo{year}{2015}).

\end{thebibliography}

    \newpage
    \noindent\textbf{Methods}
\\

    \noindent\textbf{Sample and device fabrication}\\
    The bilayers used in this study were grown by magnetron sputtering. A 15 nm Py layer was grown on the (001)-oriented PMN-PT substrate (0.5 mm thick) at room temperature, followed a SiO$_x$ capping layer of 4 nm to protect the films from oxidation. The nonlocal devices, the samples were first spin coated with photoresist, then the patterns  were made by photolithography followed by film deposition. The length of the fabricated Py strips is 50 $\mu$m, and the width is 1.5 $\mu$m.
    %The PSTO film in this study 
    %The permalloy strips in this study
    %The BFO used in this study were grown by  pulsed laser deposition. A 100 nm BFO layer was grown at a substrate temperature 800$^\circ$C and oxygen partial pressure 0.015 mbar. After the growth, the sample was cooled to room temperature in an oxygen environment of 300 mbar at a cooling rate of 20$^\circ$C/min. The devices were fabricatd using photolithography. Afterwards, the Pt layer of 10 nm were deposited by magnetron sputtering.  
    \\
\\
    %\noindent\textbf{Ferroelectric polarization switching}\\
    %The ferroelectric polarizations were switched using a ferroelectric tester (Precision Multiferroic, Radiant Technologies), with a fixed frequency of $10 kHz$ under room temperature. Two probes were placed on the Pt layer. \\
\\
    {\noindent\textbf{Theory of ferron injection, detection, and transport}
\\
An applied electric field $\mathbf{E}$ in a metallic ferromagnet with magnetization \(\mathbf{m}\) generates transverse charge and spin currents along $\mathbf{m}\times\mathbf{E}$ by the anomalous and spin Hall effects, respectively, leading to charge and spin accumulations near the edges. The transverse electric and spin currents reads
\begin{align}
\mathbf{J}_{H}^{c}=\frac{(\sigma_{\uparrow}+\sigma_{\downarrow})}{e}\boldsymbol{\nabla}\mu_{c}+\frac{\sigma_{\uparrow}-\sigma_{\downarrow}}{e}\boldsymbol{\nabla}\mu_{s}+\sigma_{\text{AHE}}\mathbf{m}\times\mathbf{\mathbf{E}}\\
\mathbf{J}_{H}^{s}=-\frac{\hbar}{2e}\left(\frac{(\sigma_{\uparrow}-\sigma_{\downarrow})}{e}\boldsymbol{\nabla}\mu_{c}+\frac{\sigma_{\uparrow}+\sigma_{\downarrow}}{e}\boldsymbol{\nabla}\mu_{s}+\sigma_{\text{SHE}}\mathbf{m}\times\mathbf{E}\right)\otimes \mathbf{m}
\end{align}
where $-e<0$ is the electron charge, $\mu_{c}=(\mu_{\uparrow}+\mu_{\downarrow})/2$ and $\mu_{s}=(\mu_{\uparrow}-\mu_{\downarrow})/2$ are the induced electrochemical potential and spin-dependent chemical potential in the transverse direction, respectively. $\sigma_{\uparrow}$ and $\sigma_{\downarrow}$ are the electric conductivities of spin-up and spin-down electrons and $\sigma_{\text{AHE}}$  ($\sigma_{\text{SHE}}$) is the anomalous Hall (spin) conductivity. With open circuit boundary condition $\mathbf{J}_{H}^{c}=0$ we can eliminate $\mu_{c}$ and 
\begin{align}
\boldsymbol{\nabla}\mu_{c}=-\alpha_{P}\boldsymbol{\nabla}\mu_{s} -\frac{e\sigma_{\text{AHE}}}{\sigma_{\uparrow}+\sigma_{\downarrow}}\mathbf{m}\times \mathbf{\mathbf{E}}\\
\mathbf{J}_{H}^{s}=-\frac{\hbar}{2e}\left\{\frac{\sigma_{\uparrow}+\sigma_{\downarrow}}{e}(1-\alpha_{P}^{2})\boldsymbol{\nabla}\mu_{s}+\sigma_{\text{SHE}}^{\prime}\mathbf{m}\times\mathbf{E}\right\}\otimes\mathbf{m}\label{Js}
\end{align}
where $\alpha_{P}=(\sigma_{\uparrow}-\sigma_{\downarrow})/(\sigma_{\uparrow}+\sigma_{\downarrow})$ is the current polarization in Py and $\sigma_{\text{SHE}}^{\prime}=(\sigma_{\text{SHE}}-\alpha_{P}\sigma_{\text{AHE}})$ an effective spin Hall conductivity.  Both the non-equilibrium spin accumulation (first term) and the anomalous Hall effect (second term) contribute to $\boldsymbol{\nabla}\mu_{c}$. The latter comes from the electric field produced by the AHE charge accumulation at the edges which for a constant drive is not time dependent due to charge conservation. The  spin-independent chemical potential gradient $-\alpha_{P}\boldsymbol{\nabla}\mu_{s}$ does not vanish when the spin filtering $\alpha_{P}\neq 0$. Since spin is not conserved, the spin accumulation fluctuates and \(\mu_{s}\) is an effective non-equilibrium chemical potential of its thermal Bose-Einstein distribution function. The in-plane components of $-\alpha_{P}\boldsymbol{\nabla}\mu_{s}$ can be interpreted as a stochastic electric field $\boldsymbol{\mathcal{E}}=-(\mathcal{\alpha_{P}}/e)\boldsymbol{\nabla}_{\perp}\mu_{s}$ that drives the spin diffusion back-current in the ferromagnet as well as the ferroelectric substrate. Here $\boldsymbol{\nabla}_{\perp}$ represents the in-plane components of the gradient operator that excites the ferroelectric order by the dipolar Stark interaction $\mathcal{H}_{int}=-\boldsymbol{\mathcal{E}}\cdot\mathbf{P}$. By solving $\mu_{s}$ with boundary conditions for the spin current Eq.~(\ref{Js}), the integrated field in the experimental configuration becomes
\begin{align}
\bar{\boldsymbol{\mathcal{E}}}=-\frac{2\lambda\theta_{\text{SHE}}^{\prime} }{w}\frac{m_{z}\alpha_{P} E}{1-\alpha_{P}^{2}} \tanh\frac{w}{2\lambda}\mathbf{x}
\end{align}
where $\lambda$ is the spin diffusion length in the ferromagnet, $\mathcal{N}_{F}$ is the density of state at the Fermi level, $\sigma=\sigma_{\uparrow}+\sigma_{\downarrow}$ is the total electric conductivity, $w$ is the width of Py strip, and $\theta^{\prime}_{\text{SHE}}=\sigma_{\text{SHE}}^{\prime}/\sigma$ the  spin Hall angle modified by the anomalous Hall effect.

Analogous to the magnon injection by the fluctuating exchange field generated by the spin accumulation, the effective electric field excites  a non-equilibrium electric polarization \(\delta P_x=\chi_{F}\bar{\mathcal{E}_x}\)  in the ferroelectric in proportion to the  ferroelectric susceptibility $\chi_{F}$. By writing  $
\delta P_x \equiv I_{p}\mu_{f}^{0} $ we introduce $\mu_{f}^{0}$, the non-equilibrium ferron chemical potential directly beneath the injector, where $I_{p}$ is a proportionality factor that depends on the non-universal ferron dipole moment. The gradient of the chemical potential drives a diffuse ferron current to the detector. With the boundary condition of a vanishing ferron accumulation at an infinite distance from the injector, i.e., $\mu_{f}(x\rightarrow{\infty})=0$, the solution of the ferron diffusion equation, $\partial_{x}^{2}\mu_{f}=\lambda_{f}^{-2}\mu_{f}$, where $\lambda_{f}$ is the ferron diffusion length, reads for \(x/\lambda_{f} \gg 1\)
\begin{equation}
\mu_{f}(x)=\mu_{f}^{0}e^{-x/\lambda_{f}}.
\end{equation}
The detection process is reciprocal to the injection: The ferron accumulation emits a fluctuating electric field \(\mathcal{E}^{\prime}\) that generates a spin accumulation gradient  $ \partial_x \mu_s (d) \propto \mathcal{E}_x^{\prime}$. The inverse spin Hall effect transforms the spin current into the voltage $V_{nl}\propto \mathcal{E}^\prime _x   (\mathbf{m}\times \hat{\mathbf{x}})\cdot \hat{\mathbf{y}}$ 
\begin{align}
V_{nl}= \mathcal{C}\frac{\lambda\theta_{\text{SHE}}^{\prime} L}{ w^2 t\sigma}\tanh\frac{w}{2\lambda}\frac{\alpha_{P}}{1-\alpha_{P}^{2}} e^{-d/\lambda_{f}} m_{z}^{2} I\propto m_{z}^2 I
\label{V}
\end{align}
where $I=w t \sigma E$ is the electric current in the injector, $t$ (\(L\)) is the thickness (length) of the contacts, and $\mathcal{C}$ is a dimensionless detector efficiency. $V_{nl}\propto m_{z}^2 I$ is the leading term of Eq.~(1) in the main text that explains the observed linear-response signal. The cubic term that senses the sign of the ferroelectric order \(P_x\)  is constrained by symmetry. The current-odd nonlinear terms in $V_{nl}$ must be even (odd) under the mirror operation  $\hat{\mathcal{M}}_{x(y)}$ with respect to the plane normal to the $x(y)$ axis 
\begin{equation}
\hat{\mathcal{M}}_{x}V_{nl}(I, m_{z},P_x)=V_{nl}(I, m_{z},\hat{\mathcal{M}}_{x}P_x)=V_{nl}(I, m_{z},P_x) 
\end{equation}
\begin{equation}\hat{\mathcal{M}}_{y}V_{nl}(I, m_{z},P_x)=V_{nl}(-I, -m_{z},\hat{\mathcal{M}}_{y}P_x)=-V_{nl}(I, m_{z},P_x). 
\end{equation}
The leading current-odd term allowed by the symmetry is therefore
\begin{equation}
V_{nl}^{(3)}\propto m_{z}P_{x}I^3. \label{V3}
\end{equation}
Combining Eq.~(\ref{V}) and Eq.~(\ref{V3}) leads to Eq. (1) in the main text
\begin{equation}
V_{nl}\approx A m_{z}^2 I+ B m_{z}P_{x} I^3 \label{Vnl}
\end{equation}
where $B$ is a constant independent of the magnetization direction and $P_{x}$. The leading non-linear response therefore vanishes with the dynamic magnetoelectric coupling when the magnetization lies in the plane ($m_{z}= 0$). The cubic term depends on the orientation of both the magnetic and electric polarizations, similar to other nonlinear transport phenomena such as the unidirectional spin Hall magnetoresistance\upcite{avci2015}.

    \noindent\textbf{Data availability}\\
    The data that support the findings of this study are available from the corresponding authors on reasonable request.\\
    
    \noindent\textbf{Acknowledgements}\\
    We are grateful for the fruitful discussions with Dr. Yuhang Li.\\    
    
    \noindent\textbf{Author contributions}\\
    X.C. conceived the experiments and supervised this study. K.S., Y.G., Y.F. and Z.G. carried out the synthesis of heterostructures and fabricated the devices. K.S. carried out the spin transport measurements. P.T. and G.B. developed the theory. Y.W., H.J., X.Z., M.W., P.H., W.S., Y.W., J.S., Q.L. and M.L. gave suggestions on the experiments. All authors discussed the results and prepared the manuscript.\\
  
    \noindent\textbf{Competing interests}\\
    The authors declare no competing interests.\\

    \noindent\textbf{Additional information}\\ Correspondence and requests for materials should be addressed to X.C.. 
    Reprints and permissions information is available at http://www.nature.com/reprints.

    \setcounter{figure}{0}
    \captionsetup[figure]{labelfont={bf}, name={Extended Data Figure}, labelsep=period}

%\newpage
\begin{figure}[t!]
    	\includegraphics[width=1\textwidth]{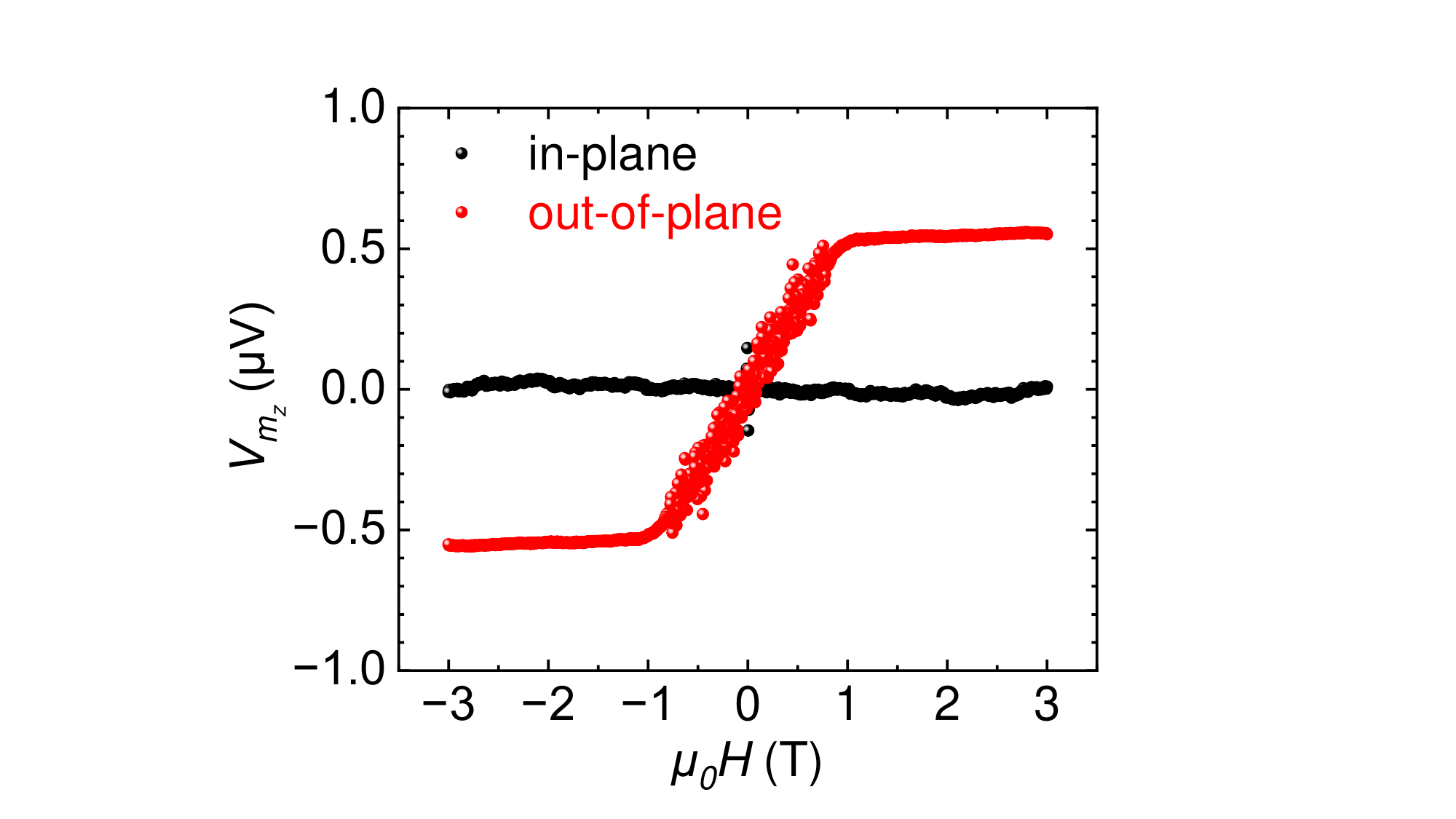}\caption{\textbf{Non-local voltage $V_{m_z}$ as a function of in-plane (black dots) and out-of-plane (red dots) external magnetic fields.}}
    	%\label{fig:Eig1}
    \end{figure}

    \begin{figure}[t!]
    	%\centering
    	\includegraphics[width=1\textwidth]{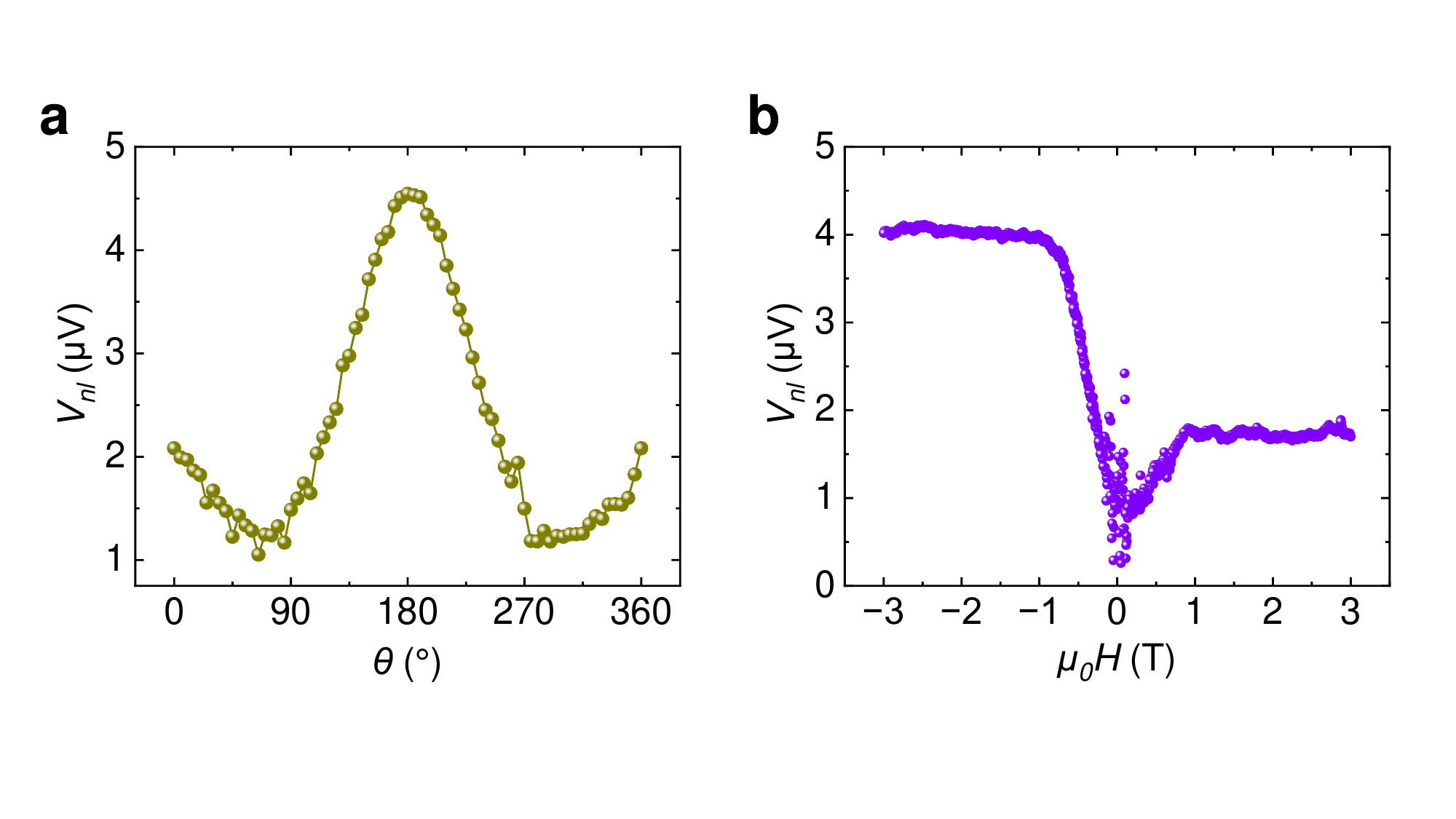}\caption{\textbf{Raw non-local voltages $V_{nl}$.} \textbf{a,} The dependence on %magnetization 
        angle $\theta$ for \(\mu_0H=3 \, \mathrm{T}\) can be fitted by $V_{nl}(\theta)=a\cos^2\theta+b\cos\theta+c$, where $a\cos^2\theta$ is the $V_{m_z^2}$ component (Figure 2\textbf{b}), and $b\cos\theta$ the $V_{m_z}$ component (Figure 2\textbf{c}).  \textbf{b,} The odd component when reversing the magnetic field from 3 Tesla to -3 Tesla %\(\theta=0^{\circ}\) to \(180^{\circ}\) 
        equals $V_{m_z}$ (Figure 3\textbf{c}), while the even one is $V_{m_z^2}$, where $\Delta{V_{m_z}}\approx-2.3 \, \mu$V and $\Delta{V_{m_z^2}} \approx1.8 \,\mu$V.}
    	%\label{fig:Eig2}
    \end{figure}

    \begin{figure}[t!]
    	
        %\centering
    	\includegraphics[width=1\textwidth]{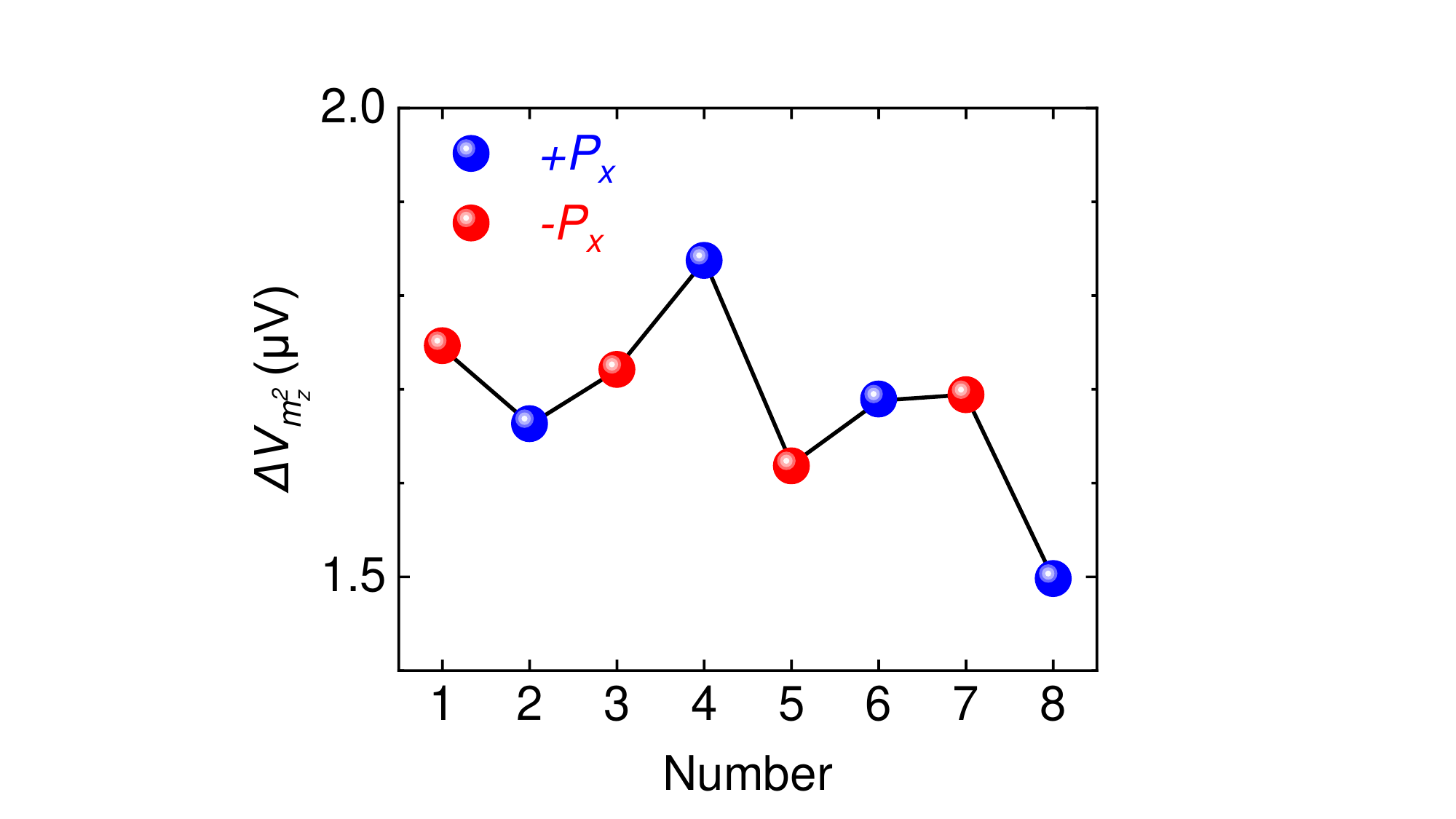}
            \caption{\textbf{$\Delta{V_{m_z^2}}$ does not change systematically under electric polarization reversal.} $\Delta{V_{m_z^2}}$ is the difference of $V_{m_z^2}$ under out-of-plane and in-plane magnetic field $H$, i.e.
            $V_{m_z^2}(\theta=0^{\circ})$ minus   $V_{m_z^2}(\theta=90^{\circ})$. 
            The data points (dots) are around 1.6 $\mu$V during the four cycles of measurements. 
            }
    	%\label{fig:Eig3}
    \end{figure}

    \begin{figure}[t!]
    	%\centering
    	\includegraphics[width=1\textwidth]{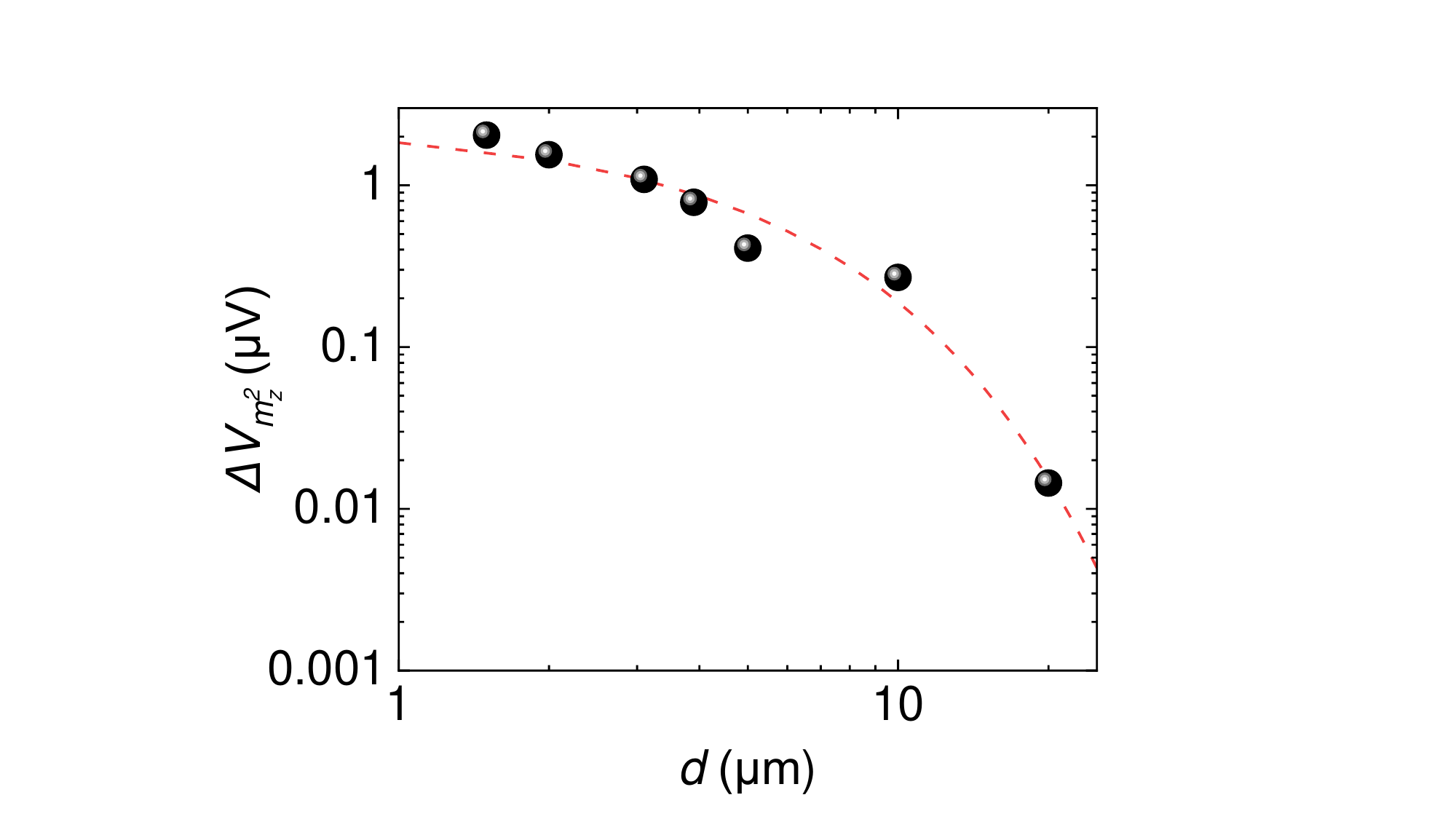}
            \caption{\textbf{Dependence of the nonlocal signal $V_{m_z^2}$ on the injector-detector spacing $d$.} Data points (dots) fit the diffusive transport model (red curve).
            }
    	%\label{fig:Eig4}
    \end{figure}

    \begin{figure}[t!]
    	%\centering
    	\includegraphics[width=1\textwidth]{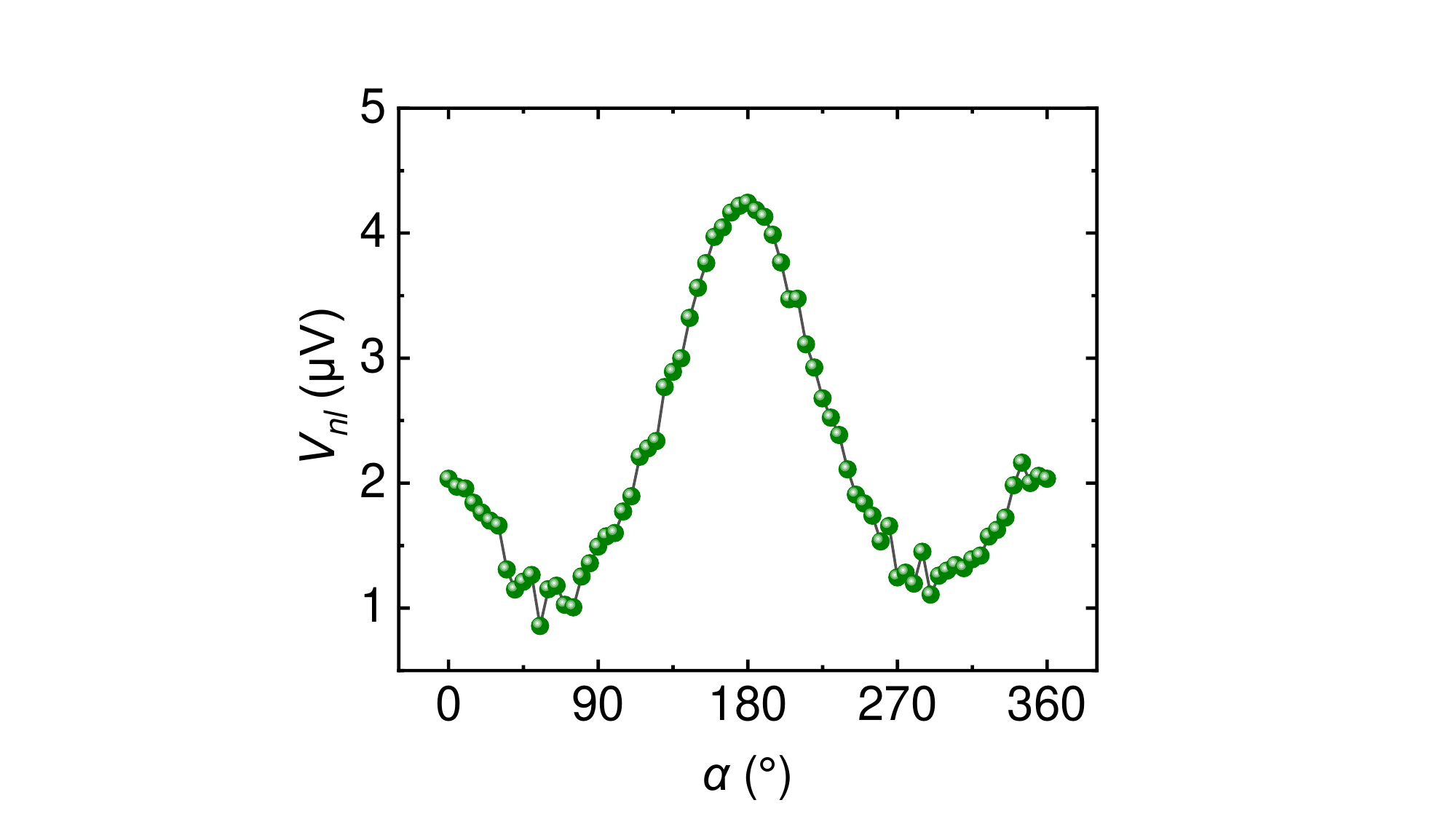}
            \caption{\textbf{$\alpha$ dependence of the raw nonlocal signal $V_{nl}$. on the injector-detector spacing $d$.} $\alpha$ represents the angle between the magnetic field $H$ and $z$-axis while $H$ rotates in the $yz$-plane.
            }
    	%\label{fig:Eig4}
    \end{figure}

\end{document}